# Laser Actuated Presentation System

[1]Atul Chowdhary, [2]Vivek Agrawal, [3]Subhajit Karmakar, [4]Sandip Sarkar

*Microelectronics Division, Saha Institute of Nuclear Physics*
*1/AF Bidhannagar, Kolkata-700064,*
*West Bengal, India*

[1]atulchowdhary@gmail.com  [2]vivek8705@gmail.com
[3]*subhajit.karmakar@saha.ac.in*  [4] *sandip.sarkar@saha.ac.in*

*Abstract*—**We present here a pattern sensitive PowerPoint presentation scheme. The presentation is actuated by simple patterns drawn on the presentation screen by a laser pointer. A specific pattern corresponds to a particular command required to operate the presentation. Laser spot on the screen is captured by a RGB webcam with a red filter mounted, and its location is identified at the blue layer of each captured frame by estimating the mean position of the pixels whose intensity is above a given threshold value. Measured Reliability, Accuracy and Latency of our system are 90%, 10 pixels (in the worst case) and 38 ms respectively.**

## I. INTRODUCTION

PowerPoint Presentations are a way of attracting audience towards your views and arguments. It is one of the most helping factors behind success of every meeting. There are various uses of power point presentations, some of them are integrated. The most popular uses of power point presentations are in modern days learning, corporate training sessions, business and marketing meetings, and sales gatherings.

Some models have already been developed to control presentation, such as a light/electronic pens, laser pointer that uses buttons with an infrared port to send commands to a PC , touch sensitive panel, some other kind of wireless handheld input devices (WHID), as well as speech and hand gesture recognition [7,8].

A laser pointer can be very useful within a classroom or business environment. Instead of pointing to the projection or LCD screen with your index finger, which gets tiring especially if the screen is large, you can point to your power-point presentations standing in one location with the laser pointer. When displaying a computer-based presentation, the user have to repeatedly return to the computer in order to press the keys that transit slides and other presentation related keyboard or mouse operations (use of function key, right click, left click to play animation, drag event to select highlighter, etc.). Ideally, presenters should be able to interact with screens from a distance, enabling them to point at the screen and control their presentation without even touching the computer or mouse. This can be achieved with our Laser Actuated Presentation System (L.A.P.S.).

## II. PREVIOUS WORKS OR RELATED WORKS

Kirstein and Muller presented a system that uses a laser pointer as a pointing device but the system suffers from a severe drawbacks. Their system may cause false triggering in case of dynamic background of the screen. Again they reported that the reliability of their system is only 50% i.e. their system is able to detect laser spot in only 50% of the frames.

A simple performance test of the laser pointer as an input device was presented in the Pebbles project [7] but they clearly concluded that the laser pointer performed the worst of any pointing device in their experiment. The authors claim that the laser pointer is not suited for selecting targets such as buttons or menus. Moreover using laser pointer, the error rate and the movement time reported by them are 6.26% and 930 ms respectively.

An inexpensive Human-computer interaction technique is introduced by Olsen [4]. In this technique a set of window events, for the laser pointer such as laser-on/off and laser-move/dwell, have been introduced. The main drawback of their system is increment in the computation time if a user moves a laser pointer so fast that the next laser spot is not found within ROI.

Winograd and Guimbretiere [5] propose a new kind of interaction techniques for large displays, which are based on "gesture and sweep" paradigm instead of the usual "point and click".

MacKenzie's, in his experiments [6], used two pointing devices (Gyropoint and Remotepoint) for remote pointing but we can see that the error rate of his experiment is much higher. According to different users, the main technical problem in their experiment was the button. It requires huge amount of force to press the button and generate any meaning command. Again, the error rate of the system is increased by significant amount as sometimes participants involuntarily performed repeated button presses on the same target.

## III. OVERVIEW

This study investigates another portable and cost-effective approach for controlling presentations. It only uses a red laser pointer (Output Power: ~5mW and Output

Wavelength: 650 nm) with an on/off button, a USB camera (fitted with a red filter) to capture the movement of the laser spot projected on the screen, a projector, a presentation screen and a presenter's computer. The presenter can use the laser pointer to control projected presentation by moving the laser spot on the presentation screen and turning the beam on or off. Fig. 1 shows how a presenter can use our system to control a slideshow application. He or she can control the slideshow using the laser pointer, avoiding the necessity of returning to the computer, which may be positioned far away. The rest of this paper will go through detail implementation of the above method.

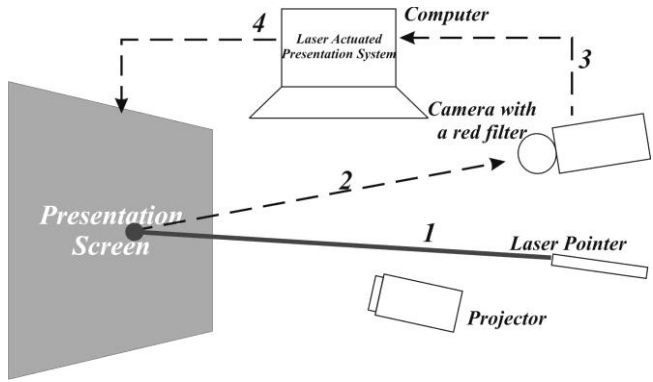

Fig. 1 Overview of the system
(1)Presenter points the laser on the screen. (2)Camera fitted with a red filter captures the images of the presentation screen. (3)Captured images are sent to the computer by the camera. (4)Computer processes the information sent by camera and executes different commands.

## IV. LASER ACTUATED PRESENTATION SYSTEM

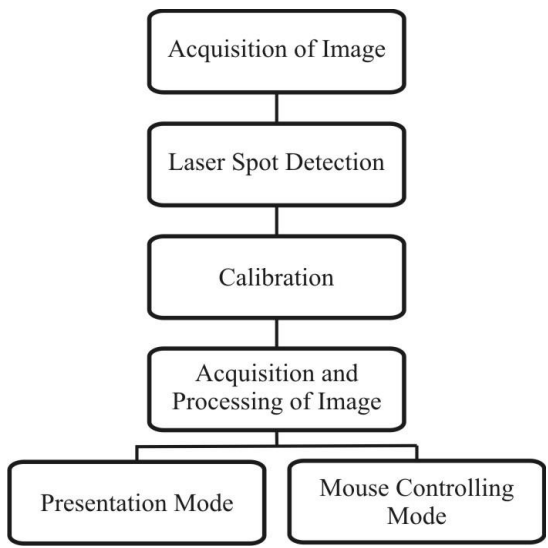

Fig. 2 Steps of the system

### A. Acquisition of Image

Using a webcam, which is able to capture frames of resolution 640 X 480 in RGB mode, frames are captured through manual triggering. In front of the webcam a red filter is mounted. Here a standard 8 bit resolution was used for all the 3 layers(R, G, and B) in order to provide intensity measures ranging from 0 to 255.

### B. Laser Spot Detection

The laser spot in the acquired image is identified by assuming the laser spot to be the brightest spot in the image. A large number of images were captured with the red filter fitted camera. These images include a set of red colored images (Fig. 3), blue colored images (Fig. 4), green colored images (Fig. 5), multiple colored images (Fig. 6), as well as different presentation images (Fig. 7). A red colored image was composed of 12 blocks (256 X 256 pixels), out of which 11 blocks having different red color intensity ranging from 0-255 and 1 white block. These blocks were randomly filled with red color and randomly positioned in the image. Similarly, Blue colored images and green colored images are generated. Multiple colored images also composed of 12 blocks. Each of these blocks was filled with randomly selected intensity value of RGB.

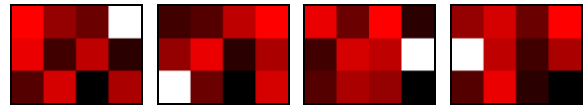

Fig. 3 Red color images.

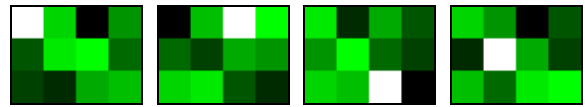

Fig. 4 Green color images.

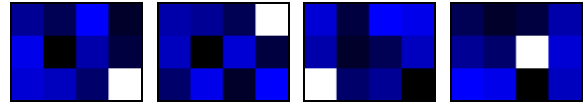

Fig. 5 Blue color images.

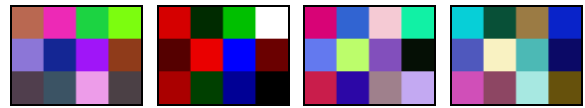

Fig. 6 Multi color images.

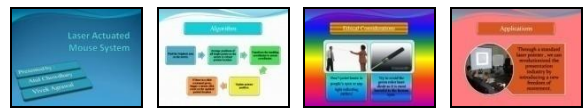

Fig. 7 Presentation images

Images were captured while the laser spot was moving on the presentation screen with the above shown images in the presentation. Observing these captured images, laser spot is detected manually to get the co-ordinates of its region. The minimum intensity value of this particular region, for each R, G, and B layer is found out and this selected value ($R_{thres}$, $G_{thres}$, $B_{thres}$) is the threshold value for each layer. This threshold value is applied on the corresponding layer of each image to find out the co-ordinates of the pixels having intensity values greater than the threshold value.

For the red layer, it is observed that apart from the laser spot there are many other pixels present having values greater than the threshold value, implying that using this threshold value we cannot distinctly identify the laser spot. Similar results are observed in case of green layer. But in case of blue layer, it is observed that the region having pixels intensity greater than the threshold value is the region of the laser spot.

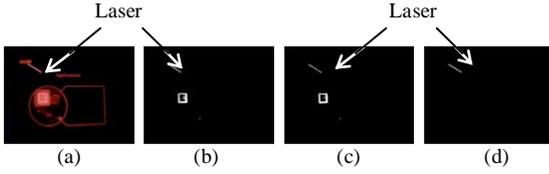

Fig. 8 Camera captured image
(a) Original Image as captured by the camera (b) Image after applying $R_{thres}$
(c) Image after applying $G_{thres}$ (d) Image after applying $B_{thres}$

Thus, for the detection of the laser spot only the blue frame of the acquired image is considered for processing as laser spot can be uniquely identified in the blue frame using the threshold mechanism.

Experimentally, (see Fig. 19) it has been found in the acquired images where the pixel intensity values are greater than 200 (i.e. threshold value) represents the bounding region of the laser spot. The mean of the bounding region is the center of the laser spot.

The co-ordinates (x, y) i.e. center of the laser spot is calculated using the following equation

$$(x, y) = \left( \frac{\sum_{i=1}^{n} x_i}{n}, \frac{\sum_{i=1}^{n} y_i}{n} \right)$$

where n represents the number of pixels in the bounding region and ($x_i$, $y_i$) represents each individual pixel values.

### C. Calibration

Unless the projector and the webcam are precisely aligned to the presentation screen, the resulting image suffers from perspective (keystone) distortions as shown in Fig. 9(a). Although some modern LCD projectors offer a form of digital keystone correction but distortion present in the image captured by the webcam can only rectified manually.

A calibration process is needed at the starting point of the system to map the detected positions of laser spot in the camera image to the corresponding positions and shapes on the screen. We employ a simple solution that uses a perspective transformation.

Our method for calibration is summarized as follows. (1) Determine the four corners of the presentation window on the screen. The user points to four corners of the screen one by one with a laser pointer for 2 seconds (10 frames) for every corner. (2) The system measures mean position of the laser spots in the camera image for each corner. (3) After the four corner positions are given, a quadrilateral is formed using these coordinates. This quadrilateral is our region of interest (ROI). (4) A transformation matrix is constructed using projective transformation method. This transformation matrix takes care of not only the perspective transformation, but also distortion introduced by rotation of the camera; thus by simply multiplying a camera point by this matrix, the computer screen coordinates can easily be determined. (5) Whenever a laser spot is detected by the camera it first checks whether it lies in our ROI, and if it does, it transforms the absolute coordinate captured by the webcam to relative coordinate in the computer screen. This also negates the offset error present between the mouse cursor and the laser spot.

Let us suppose the actual screen resolution of the computer is 1024 x 768 (Though the system works for any screen resolution, in this paper the calculations are shown based on 1024 x 768 screen resolutions). Now the Fig. 9(a) denotes the camera 8 projection screen while the dark part Fig. 9(b) is the Region of interest i.e. it is the actual projection of the computer. So we need to map the whole quadrilateral into a rectangle of 1024 x 768 pixels resolution. This mapping is done using the projective transformation technique [9].

In projective transformation, using the co-ordinates of four corners A, B, C, D as obtained from the camera in Fig. 9(a) and $A'$, $B'$, $C'$, $D'$ as obtained from the screen resolution of the computer in Fig. 9(b), a transformation matrix P(3X3) is constructed.

$$P = \begin{bmatrix} P_1 & P_2 & P_3 \\ P_4 & P_5 & P_6 \\ P_7 & P_8 & P_9 \end{bmatrix}$$

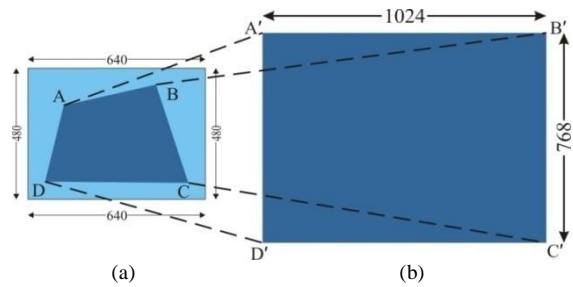

Fig. 9 Calibration screen

Using the equation given below the camera co-ordinate (x, y) is mapped into absolute co-ordinate (X, Y) i.e. computer screen co-ordinates.

$$\begin{bmatrix} X' \\ Y' \\ W' \end{bmatrix} = \begin{bmatrix} P_1 & P_2 & P_3 \\ P_4 & P_5 & P_6 \\ P_7 & P_8 & P_9 \end{bmatrix} \begin{bmatrix} X \\ Y \\ 1 \end{bmatrix}$$

$$\begin{bmatrix} X \\ Y \end{bmatrix} = \begin{bmatrix} X'/W' \\ Y'/W' \end{bmatrix}$$

### D. Processing of acquired image

The acquired frame is divided into three regions i.e. lower region, middle region, upper region, Fig. 10. From the acquired frame, if the co-ordinate of the mean position of the laser is found to be present in the lower region, based on its location presentation mode or mouse controlling mode will be activated.

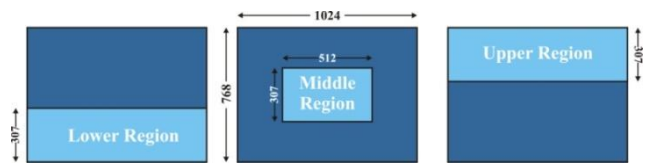

Fig. 10 Regions of the image

## E. Presentation mode:

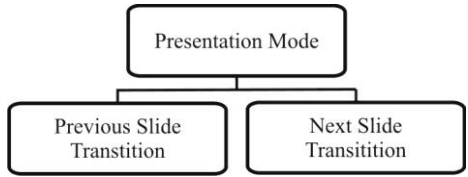

Fig. 11 Presentation Mode

In the presentation mode, the laser spot will be first searched in the lower region of the acquired frames, if it is present there then it will be searched in the middle region and if it is found over there, it will be searched in the upper region of the next acquired frames. Based on the position of the laser spot in these regions slide transitions will take place.

*1) Next slide transition:* For a next slide transition to take place, laser spot must appear in the left side of the lower region, Fig. 12(a). On detecting the laser spot in the above said region, within the next 5 frames, it should appear in the middle region, Fig. 12(b), and after that within the next 10 frames it should appear in the right side of the upper region, Fig. 12(c). When all the three conditions mentioned above are satisfied, there will be next slide transition in the presentation, Fig. 12(d).

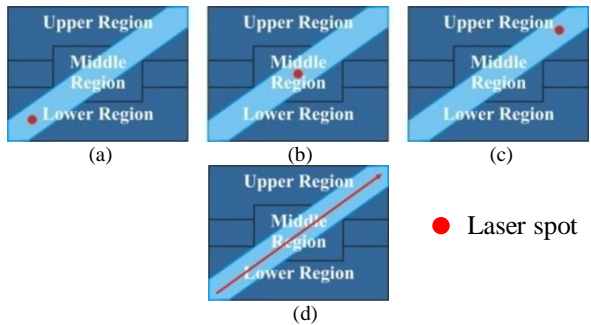

Fig. 12 Steps involved in Next slide transition

*2) Previous slide transition:* For a previous slide transition to take place the laser spot must appear in the right side of the lower region, Fig. 13(a). On detecting the laser spot in the above said region, within the next 5 frames, it should appear in the middle region, Fig. 13(b), and after that within the next 10 frames it should appear in the left side of the upper region, Fig. 13(c). When all the three conditions mentioned above are satisfied, there will be previous slide transition in the presentation, Fig. 13(d)

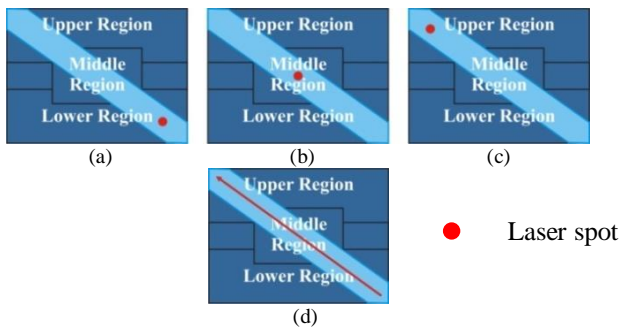

Fig. 13 Steps involved in previous slide transition

## F. Mouse controlling mode:

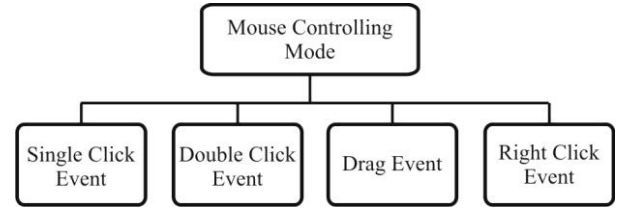

Fig. 14 Mouse Controlling Mode

For the activation of mouse controlling mode, the laser spot must appear in the middle of the lower region, Fig. 15(a). On detecting laser spot in the above mentioned region, within the next 10 frames captured, it should appear in the upper region, Fig. 15(b). If the above mentioned conditions are satisfied, mouse controlling mode will be activated, Fig. 15(c).

The mouse controlling mode will be automatically deactivated once the laser spot is not detected continuously for 5 frames captured.

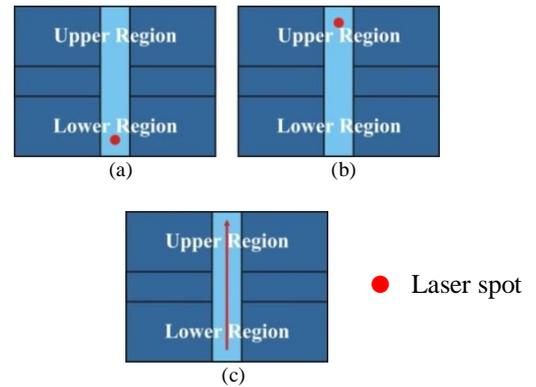

Fig. 15 Steps involved in the activation of Mouse controlling mode

In this mode, the path of the mouse cursor will be governed by the laser light, the cursor on the screen will move according to laser spot.

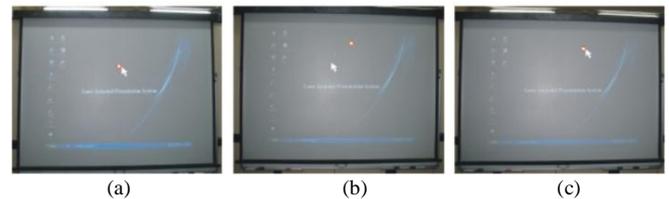

Fig. 16 Cursor and laser pointer interaction
(a) Before a laser point moves (b) Laser point moves (c) Cursor chases it.

*1) Single click event:* In any presentation, single click event is mainly required to click any hyperlink, animation, play and pause of video. Single click event will be triggered if for any successive 5 frames, the laser spot positions remain in the vicinity of the target which is about 2% of the screen resolution. After the completion of the single click event the system goes back to the normal mouse controlling mode.

*2) Double click:* Similar to the Single click event Double click event will be triggered if the laser spot stays on the target for successive 10 frames. After the completion of the double click event the system returns to the normal mouse controlling mode.

*3) Drag and Drop event:* In any presentation, drag event is mainly required to use the highlighter or pen option. Drag event will be triggered if the movement of the laser spot will be in the vertical up and down direction as shown in Table II. The laser spot should be moved in such a way that it traverses a distance of more than 20% of the screen resolution along y-axis with a tolerance level of less than 10% of the screen resolution along x-axis. This movement enables the Drag and Drop event. The info box in Fig. 18(f) shows Drag and Drop event has been enabled. Once the event is enabled the user can start dragging an object if for any successive 5 frames captured, the laser spot positions remain in the vicinity of 2% of the screen resolution in each direction at a particular point. The drag event will get activated on the position were laser spot remained for 5 successive frames. This action is equivalent to the press of the left mouse button. To drop the selected item anywhere on the screen, we again need to hold our laser pointer in the vicinity (2% of the screen resolution) of that particular position (where we want to drop the selected item) for 5 frames. This action is equivalent to the release of the left mouse button. After the completion of the drag and drop event, the system goes back to the mouse controlling mode. To disable the drag event at any moment, we just need to switch off our laser pointer.

*4) Right click event:* Right click event will be triggered if the movement of the laser spot will be in the horizontal right and left direction as shown in Table II. The laser spot should be moved in such a way that it traverses a distance of more than 20% of the screen resolution along x-axis with a tolerance level of less than 10% of the screen resolution along y-axis. The info box in Fig. 18(e) shows right event has been enabled. To activate the right click event at any position we must hold our laser pointer in the vicinity (2% of the screen resolution) of that particular position for 5 frames. This action is equivalent to the press and release of the right mouse button. After the completion of the right click, the system goes back to the mouse controlling mode. To disable the right click event at any moment, we just need to switch off our laser pointer.

TABLE I
COMMANDS FOR COMPUTER BASED PRESENTATION

| Function | Drawings | Motion of Laser Pointer |
|---|---|---|
| Next Slide | ↗ | From Lower Left corner to Upper Right Corner |
| Previous Slide | ↖ | From Lower Right corner to Upper Left Corner |

TABLE II
MOUSE FUNCTION REPLACEMENT

| Mouse Function | Alternative Command | Drawings |
|---|---|---|
| Single click | Hold at a particular spot for 5 frames | ● |
| Double Click | Hold at a particular spot for 10 frames | ● |
| Drag and Drop | Move the pointer in up and down direction | 1↑↓2 |
| Right Click | Move the pointer to and fro direction | 1→ ←2 |

V. IMPLEMENTATION

The system is very user friendly from both audience as well as presenter's point of view. From audience point of view, this system is very friendly as it avoids the cursor echo problem. A typical laser point detection system checks whether the laser point is located on the screen and, if it is on, its relative position to the screen in the captured image is transformed into a cursor coordinate. Thereby audience cannot see a laser point moving on the screen without a coupled cursor motion. This phenomenon is called cursor echo that denotes visual feedback represented by a cursor due to a laser pointer action in the system [4]. But in our system, the laser point gets coupled to the cursor only when the presenter wants to use the laser for mouse controlling.

The system makes use of an efficient GUI (created using javax.swing.* class) in form of a small rectangular box which is located at the top left corner of the screen (the box automatically moves to the right top corner if the presenter focuses the laser on the top left corner). From presenter's point of view, this GUI is helpful as it states the current status of the system. In short, this GUI acts as a guide to the presenter.

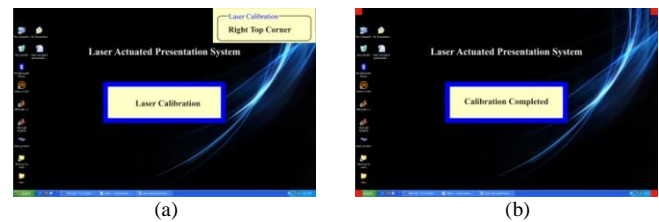

Fig. 17 GUI of Calibration Screen
(a)GUI instructing the presenter to point the laser at the right top corner through the info box at right top corner (b)GUI showing the completion of calibration process and the four red boxes indicating the corners of the screen.

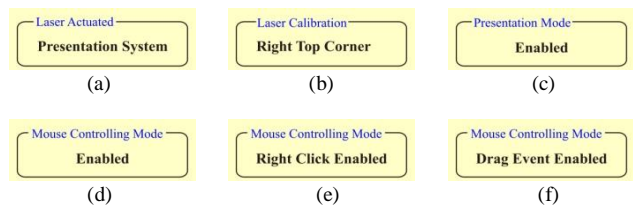

Fig. 18 GUI of different info box

(a)Info box indicating the application is active (b) Info box asking the user to point to the right top corner of the screen for calibration (c) Info box indicating presentation mode is active (d) Info box indicating mouse controlling mode is active (e) Info box indicating right click event has been triggered. (f) Info box indicating drag and drop event has been triggered

The system used is quite flexible, allowing for a large range of display size and resolution. The system evaluated here was composed of a 30 fps RGB video camera with a resolution of 640×480 pixels fitted with a red filter (Cokin filter), an XGA (1024×768 pixels) video projector (Optoma EP719 DLP Portable Projector 2000). Both the camera and the projector were connected to an Intel Core Duo T2300E processor Laptop with 512 MB DDR2 RAM, running at 1.66 GHz. The application has been designed using MATLAB 7.1. When running, the application used a maximum of 26 % of the CPU resources.

The distance between the presentation screen and the camera was 3.8m. Distance between projector and presentation screen was 4m. The system was tested under the light conditions. The luminance of the room was measured using Minolta IVF Autometer. The shutter speed of the Autometer was adjusted to $1/10^{th}$ sec and exposure was set to 100 ISO.

All the auto-adjustments of the camera were turned off. The system can work as long as the three following conditions are fulfilled:

- The display screen is entirely visible (and in focus) inside the field of view of the camera.
- There is no relative movement between the screen and the camera once the system is started.
- The laser spot is brighter than anything else in the field of view of the camera.

## VI. PERFORMANCE RESULTS

The performance results obtained are summarized in Table III below.

TABLE III
PERFORMANCE RESULTS

| Criteria | Results |
|---|---|
| Reliability | 90 % |
| Accuracy | 10 screen pixels (worst case) |
| Latency | 26/56/38 ms (min/max/average) |

### A. Reliability

The reliability of the system came out to be 90%, meaning that the laser spot was detected in 90% frames, i.e. the system showed a very good reliability. This result has been achieved by performing real time experiment under different light conditions ranging from 1 lx. to 130 lx. The system is using the threshold concept, and the following result as shown in Histogram (fig. 19) indicates the threshold value of 200 is appropriate to detect laser spot on any frame. The system is able to detect the laser spot when the laser spot is moving within a speed range of 0-2000 pixels/second. Since laser spot is the brightest object in the field of view of the camera screen, so in case of dynamic background also, laser spot is easily detected.

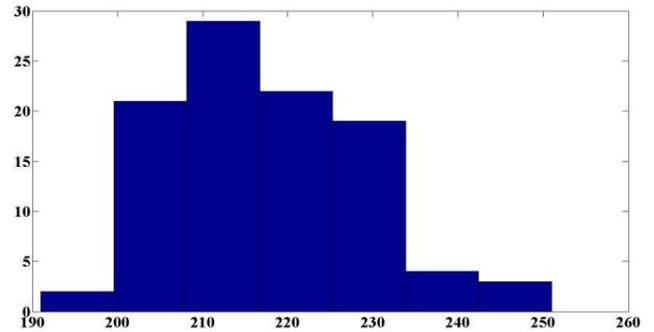

Fig. 19 Histogram of threshold values under different light conditions and color backgrounds for laser spot moving with a speed of 0-2000pixels/sec.

### B. Accuracy

Accuracy of the system defines the difference (in pixel) between the tip of the on-screen cursor pointer and the stationary laser spot at different regions of the screen. The measured accuracy was of 10 camera pixels in the worst cases, this type of similar result has been reported before [4, 10]. The main reason of getting this result is related to the camera-to-display mapping method (calibration technique) used. In fact, the analysis of the accuracy of the system revealed that the accuracy varies from one region of the display to another, with a 10 pixel discrepancy in the worst case. Radial distortion in the camera lens affects accuracy of the system; this can be verified by analyzing the inaccuracies that tend to shift towards the center and the display sides. Better accuracy can be attained if the four corners of the presentation screen are selected precisely by the laser pointer, Fig. 20(b). For this, it is recommended to point the laser on the presentation screen from the minimum possible distance.

The distance between the camera and the presentation screen also plays a vital role in determining the accuracy of the system, Fig. 20(a). If the camera is placed very far away from the screen then the ROI (Fig. 9) will be very small and mapping will not be performed precisely. In order to get best results, camera must be placed such that the field of view of camera captures only the display screen.

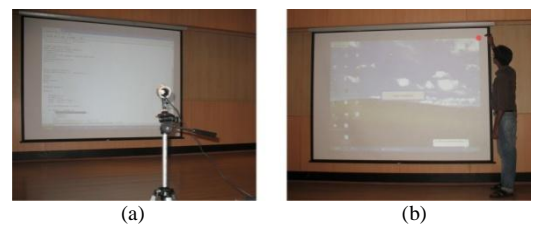

(a)        (b)

Fig. 20 Factors affecting accuracy

### C. Latency

Latency is one of the most serious issues as far as performance of the system is concerned. To measure latency,

the cursor was placed at an arbitrary position in the computer screen. The laser pointer was then activated to point at an arbitrary region within the screen area. This was recorded by a 30 fps video camera. The time taken by the cursor to reach the static laser spot was then measured from the resulting video by counting the number of frames between the two events. Several readings were taken to get the final result; final result was stated in the form of best, average, worst case in Table III.

### VII. DISCUSSION & FUTURE WORK

Reliability of the system depends on the threshold value, but there may be some lighting conditions or background color where the chosen threshold value is not appropriate to detect the laser spot. In that case reliability of the system will be decreased. To solve this problem, instead of using static threshold value (200 for this system) for all conditions, adaptive threshold value can be introduced in the system to get better results. Accuracy can be enhanced using a high resolution camera to capture image. Latency of the system can also be minimized by using faster camera; it would reduce the lag impression, introduced by user when the laser spot is moved rapidly on the screen. Latency can further be improved by embedding the algorithm on FPGA or similar hardware platforms. Although the system has been designed keeping in mind the user friendliness but there is always a scope for improvement. According to the convenience of a user some other alternatives gestures for mouse replacement functions through laser could be thought of.

With some modifications, this technique can be helpful for physically challenged people. For physically challenged people laser pointer will be mounted on an adjustable headband, allowing a hand free operation. The movement of the headgear will control the mouse movements as well as other mouse functions. To implement this scheme, we would require estimating the movement time and the index of difficulty for each command [11]. This work is under progress.

### VIII. CONCLUSION

In this paper, we have designed a human-computer interaction scheme for PowerPoint presentation. The interaction is actuated by simple patterns drawn on the presentation screen by a laser pointer. These patterns activate different basic operations of Presentation as well as mouse controlling operations like click, drag and drop etc. We could achieve Reliability of 90%, Accuracy of 10 pixels (in the worst case) and average Latency of 38 ms. We have implemented this scheme on a demo system which doesn't require any specialized hardware [2, 6]. This application can be useful for physically challenged people as it can provide hands free operation.

### ACKNOWLEDGMENT

We wish to thank, Prof. Kuntal Ghosh (Indian statistical Institution) for helpful advice on our project, Saha Institute of Nuclear Physics (SINP) for supporting this research project.